\documentclass[twocolumn]%,linenumbers]
{aastex631}
\usepackage{amsmath}
\usepackage{amssymb}
\usepackage{aas_macros}
\usepackage{appendix}
\usepackage{gensymb}

\begin{document}

\title{A Possible Common Physic Picture Reflected by the Gamma Ray Emission of the Galactic Center}

\correspondingauthor{Yi-Qing Guo}
\email{guoyq@ihep.ac.cn}
\correspondingauthor{Si-Ming Liu}
\email{liusm@swjtu.edu.cn}
%\author{**********************}
%\affiliation{*********}
\author{Lin Nie}
%\affiliation{School of Mechanics and Aerospace Engineering, Southwest Jiaotong University, Chengdu, 610031, China}
\affiliation{School of Physical Science and Technology, Southwest Jiaotong University, Chengdu, 610031, China}

\affiliation{Key Laboratory of Particle Astrophysics, Institute of High Energy Physics, Chinese Academy of Sciences, Beijing, 100049, China
}

%\author{Yu-Hai Ge}
%\affiliation{Key Laboratory of Particle Astrophysics, Institute of High Energy Physics, Chinese Academy of Sciences, Beijing, 100049, China
%}

\author{Yi-Qing Guo}
\affiliation{Key Laboratory of Particle Astrophysics, Institute of High Energy Physics, Chinese Academy of Sciences, Beijing, 100049, China
}
\affiliation{University of Chinese Academy of Sciences, Beijing 100049, China}
\affiliation{TIANFU Cosmic Ray Research Center, Chengdu 610000, China}
\author{Si-Ming Liu}
\affiliation{School of Physical Science and Technology, Southwest Jiaotong University, Chengdu, 610031, China}

\begin{abstract}
Long-term observations of the Galactic center by Fermi and HESS have revealed a novel phenomenon: the high-energy gamma-ray spectrum from the Galactic center exhibits a double power-law structure. In this study, we propose a new explanation for this phenomenon. We suggest that the low-energy (GeV) power-law spectrum originates from interactions between trapped background ``sea" cosmic ray particles and the dense gaseous environment near the Galactic center. In contrast, the bubble-like structure in the high-energy (TeV) spectrum is produced by protons accelerated during active phases of the Galactic center, through the same physical process.
Based on this framework, we first calculate the gamma-ray emission generated by cosmic ray protons accelerated in the Galactic center. Then, using a spatially-dependent cosmic ray propagation model, we compute the energy spectrum of background ``sea" cosmic ray protons and their associated diffuse gamma-ray emission in the Galactic center region. The results closely reproduce the observations from Fermi-LAT and HESS, suggesting that their long-term data support this picture: high-energy cosmic rays in the local region originate from nearby cosmic ray sources, while low-energy cosmic rays are a unified contribution from distant cosmic ray sources.
We predict that some extended Galactic sources, which remain undetectable in the GeV energy range, may become observable in the TeV range. We hope that future observations will detect more such sources, allowing us to further test and validate our model.

\end{abstract}

\keywords{Galactic Center, Cosmic ray propagation, Diffuse $\gamma$-ray emission}

\section{Introduction} \label{sec:intro}
In recent years, with the operation of experiments of higher sensitivity, especially space-based experiments such as AMS-02\citep{2015PhRvL.114q1103A}, Fermi-LAT\citep{2016ApJS..223...26A,2017PhRvL.118i1103A}, PAMELA\citep{2009Natur.458..607A,2011Sci...332...69A}, ATIC\citep{2007BRASP..71..494P,2009BRASP..73..564P}, CREAM\citep{2010ApJ...714L..89A}, and our country's dark matter satellite DAMPE\citep{2019SciA....5.3793A}, more refined measurements have been obtained in the spectra of cosmic rays and gamma rays. Precise measurements of cosmic rays have revealed ``anomalies" that cannot be anticipated by the so-called ``standard" cosmic ray propagation models.

For instance, the spectra of all primary cosmic rays harden at high energy. Data released by AMS-02 indicate that the cosmic ray spectrum deviates from a simple power law and hardens at energies exceeding several hundred GeV\citep{2015PhRvL.114q1103A,2021PhR...894....1A}, a phenomenon that has been mutually confirmed by different observational instruments\citep{2019SciA....5.3793A,2009Natur.458..607A,2011Sci...332...69A,2007BRASP..71..494P,2009BRASP..73..564P,2010ApJ...714L..89A}. Moreover, the spectra of all secondary cosmic rays also harden in this energy range, and the spectra of secondary cosmic ray particles beyond this energy range are harder than those of primary cosmic ray particles\citep{2018PhRvL.120b1101A}. The latest observations of secondary and primary cosmic ray particle fluxes and their ratios by DAMPE further confirm that the secondary-to-primary ratio of cosmic rays exceeds at high energies\citep{2016PhRvL.117i1103A,2022SciBu..67.2162D}. To uniformly explain these ``anomalous" phenomena, spatially dependent cosmic ray propagation models have been developed.
It follows three fundamental principles dominating cosmic ray propagation. 
Firstly, the propagation of cosmic rays depends on the local distribution of matter; secondly, local cosmic ray sources act like potential wells to confine the escape of local cosmic rays and like potential barriers to prevent the entry of foreign cosmic rays into their region; thirdly, cosmic rays in local regions are mainly composed of two components, namely component A from distant cosmic ray sources accelerated and diffused to the local region and another component B produced by local sources. The former dominates the distribution of cosmic rays at low energies, while the latter dominates at high energies, with local sources leading the high-energy phenomena in the Milky Way\citep{2024PhRvD.109f3001Y,2024ApJ...974..276N}. This propagation scenario well explains the aforementioned cosmic ray spectral observations. However, gamma rays are one of the best means to study the origin and propagation of cosmic rays, and observations combining the diffuse gamma rays of the Milky Way and the local radiation of local cosmic ray sources are expected to validate this propagation scenario.

The Galactic Center, due to the accretion activity of a supermassive black hole, is a widely discussed site for cosmic ray acceleration. Thanks to the observations by Fermi-LAT and HESS, the gamma-ray radiation from the Galactic Center has been well measured. Interestingly, the energy spectra obtained by Fermi-LAT\citep{2011ApJ...726...60C} and HESS\citep{2009A&A...503..817A} cannot be simply connected by a power law; the overall spectrum from the GeV to TeV energy range presents a double power-law structure, posing a challenge to understanding its radiation mechanism. It is commonly assumed that two components exist to separately explain the Fermi-LAT and HESS data\citep{2011ApJ...726...60C,2012ApJ...757L..16F,2013JPhG...40f5201G}. Some also believe that low-energy and high-energy cosmic rays in the Galactic Center undergo different diffusion processes\citep{2012ApJ...753...41L,2012ApJ...748...34K}, leading to different energy spectra in the GeV and TeV energy ranges.

In this work, based on the physical scenario of the spatially dependent cosmic ray propagation model, we propose a possible ``common" model for the diffuse gamma-ray radiation of the Galactic Center that may be universally present in cosmic ray sources in the Milky Way, studying the dual-component spectral structure characteristics currently observed from the Galactic Center. Based on the observations of the Galactic Center by Fermi-LAT and HESS, we investigate the nature of the double power-law structure of the Galactic Center's radiation spectrum, test the scenario of the spatially dependent cosmic ray propagation model we proposed, and look forward to the universality of the observed energy spectra presenting a double power-law structure around cosmic ray sources in the Milky Way. This paper is structured as follows: Section \ref{sec:method} introduces the research methods; Section \ref{sec:result} discusses our results; and Section \ref{sec:conclusion} summarizes and prospects this paper.
\begin{figure*}[t]%[htbp]
    \centering
    \includegraphics[width=0.95\linewidth]{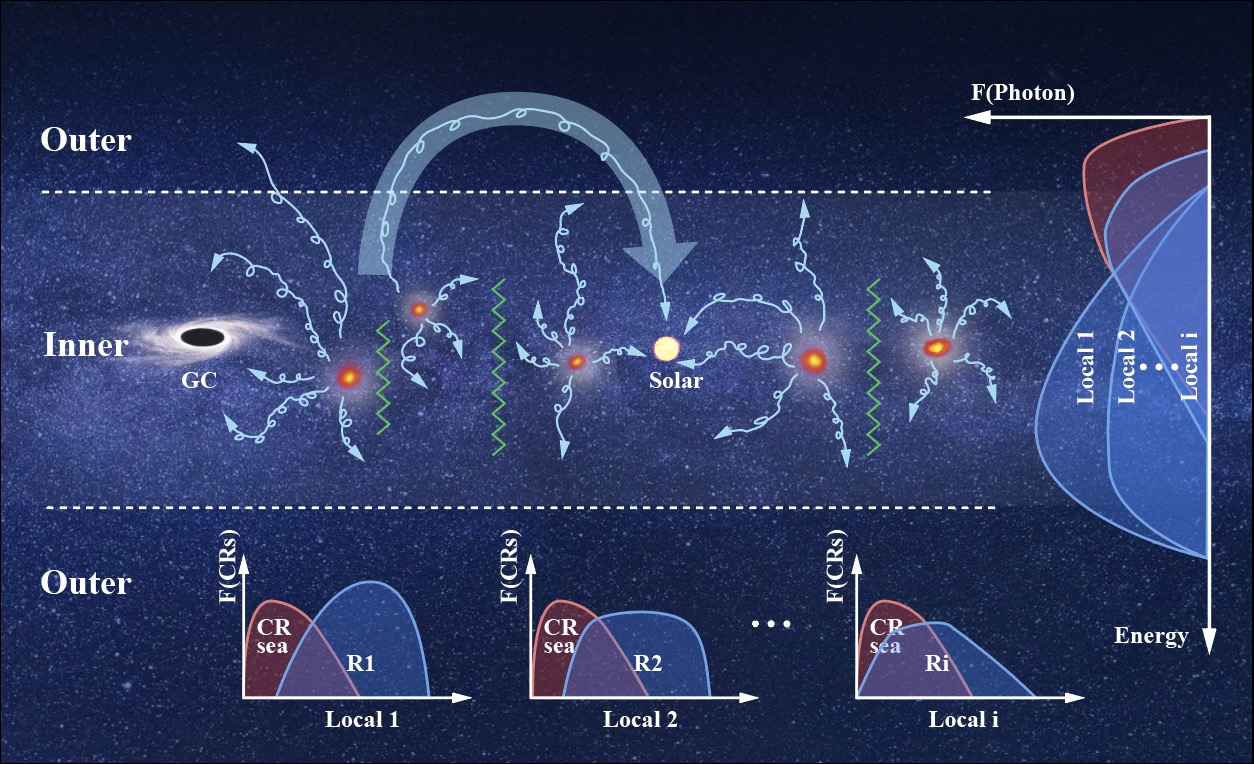}
    \caption{A schematic diagram of cosmic ray propagation in the Milky Way galaxy and the radiation produced at different locations. The diagram briefly describes the dual role of the cosmic ray halo: on the one hand, the halo acts like a potential barrier, restricting the propagation of incoming cosmic ray particles; on the other hand, it behaves like a potential well, trapping and capturing cosmic ray particles. As a result, cosmic ray particles from distant regions must propagate through the outer halo of the galaxy to reach more distant halo regions. Subsequently, these cosmic ray particles, which have propagated from the outer halo, are captured by the local halo, leading to two distinct cosmic ray components within the local halo. The distribution of gamma rays within the halo also exhibits similar characteristics.}
    \label{cartoon}
    \end{figure*}

\section{MODEL \& METHODOLOGY} \label{sec:method}
The spatially dependent propagation model, which successfully explains a variety of cosmic ray observational data from the perspective of propagation, provides an excellent viewpoint. As illustrated in Figure \ref{cartoon}, this model posits that the cosmic ray spectrum in the Milky Way has a double-component structure. The key physical point forming this double-component structure is that the cosmic ray Halo of the Milky Way not only acts like a potential well to trap and constrain the cosmic ray particles injected by sources but also like a potential barrier to prevent the passage of foreign particles. Due to the relatively fewer cosmic ray Halos in the outer halo compared to the galactic disk, particles undergo a faster diffusion process. Therefore, cosmic rays accelerated by distant sources may diffuse rapidly through the outer halo, forming a softer cosmic ray ``sea" component; whereas cosmic ray particles accelerated by local sources, due to the confinement and trapping by the cosmic ray Halo, have a longer residence time and thus possess a harder spectral index. This component, varying with different Halos or positions in the Milky Way, is a non-steady-state component. Based on this scenario, we construct a ``common" model that may be universally present in the cosmic ray Halos of the Milky Way, suggesting that there are two cosmic ray components within the Halo: one component is provided by the acceleration of point sources within the Halo, depending on the characteristics of the sources; the other component is a stable component from the cosmic ray ``sea".

\subsection{Diffusion Around the Galactic Center}
The estimation results from gamma-ray observations suggest that the diffusion coefficient around the cosmic ray Halo may be several hundred times smaller than the average value for the entire Milky Way. This slow diffusion region could be a turbulent area generated by cosmic ray particles escaping from sources. We consider the Halo as a slow diffusion region, where cosmic ray sources within the Halo accelerate cosmic rays and continuously inject them into the Halo. However, due to the smaller diffusion coefficient of cosmic ray particles in the Halo, the Halo acts like a potential well, restricting the rapid diffusion of cosmic ray particles.

Therefore, in our current work, we consider the region near the Galactic Center as a slow diffusion area. Observations indicate that although the Galactic Center is relatively quiet at present, there are still signs of weak activity. It is highly likely that high-energy cosmic ray particles are accelerated during such activities. We assume that cosmic ray protons from the Galactic Center are accelerated during its flare activities, then freely diffuse in the surrounding slow diffusion environment, and finally escape into the interstellar medium (ISM) environment.The propagation process of cosmic ray particles from the Galactic Center is described by the following propagation equation\citep{2011ApJ...726...60C,2013JPhG...40f5201G}:
\begin{equation}
\frac{\partial \phi}{\partial t}=\frac{D(E)}{r^2} \frac{\partial}{\partial r} r^2 \frac{\partial \phi}{\partial r}+\frac{\partial}{\partial E}[b(E) \phi]+N(E) \delta(t) \delta(\mathbf{r}),
\end{equation}
where $\phi(r,E,t)$ is the propagated proton flux as a function of space, energy and time, D(E) is the diffusion coefficient, N(E) is the injection spectrum of particles at t = 0 and r = 0, and $b(E) = dE/dt$ is the energy loss rate, which is important for electrons but negligible for protons.
Here, the diffusion coefficient is a spherically symmetric "two-zone" diffusion, with slow diffusion near the Galactic center and faster diffusion in the interstellar medium. Mathematically, the diffusion coefficient is described as follows \citep{2019ApJ...879...91J}:
\begin{equation}
D=\beta\left(\frac{\mathcal{R}}{\mathcal{R}_0}\right)^{\delta} \begin{cases}D_h, & r<r_i \\ D_h\left[\frac{D_0}{D_h}\right]^{\frac{r-r_i}{r_o-r_i}}, & r_i \leqslant r \leqslant r_o \\ D_0, & r>r_o,\end{cases}
\end{equation}
where $\beta$ represents the particle velocity in units of the speed of light, $D_0$ is the normalization of the diffusion coefficient in the general ISM, $D_h$ is the normalization for the diffusion coefficient within the SDZ with radius $r_i$, $R$ is the particle rigidity, and $\rm R_0=4GV$ is the normalization (reference) rigidity. The zone between $r_i$ and $r_o$ is a transition layer where the normalization of the diffusion coefficient increases exponentially from $D_h$ to the ISM value $D_0$.

The cosmic ray particles are injected into the halo by the Galactic center propagate through diffusion. A single power-law cutoff function describes the injection spectrum:
\begin{equation}
Q(E) = Q_{0}E^{\gamma} \exp (-E / E_{cut})
\end{equation}
Here, $E_{cut}$ is break energy, $\gamma$ is spectral indices for low and high energy bands, and $\rm Q_{0}$ represents the injection rate. 

\subsection{Diffusion of Background ``Sea" Cosmic Ray Particle}
We believe that the softer component of the Galactic background ``sea" cosmic rays dominates the low-energy component in the Halo (here referring to the Galactic center). Therefore, it is necessary to calculate this component using a spatially dependent cosmic ray propagation model. The propagation of Galactic cosmic rays is described by the following diffusion equation:

\begin{equation}
    \begin{aligned}
        \frac{\partial \psi(\vec{r}, p, t)}{\partial t}= & Q(\vec{r}, p, t)+\vec{\nabla} \cdot\left(D_{x x} \vec{\nabla} \psi-\vec{V}_c \psi\right) \\
        & +\frac{\partial}{\partial p}\left[p^2 D_{p p} \frac{\partial}{\partial p} \frac{\psi}{p^2}\right] \\
        & -\frac{\partial}{\partial p}\left[\dot{p} \psi-\frac{p}{3}\left(\vec{\nabla} \cdot \vec{V}_c\right) \psi\right]-\frac{\psi}{\tau_f}-\frac{\psi}{\tau_r}
        \end{aligned}
        \label{eq3}
\end{equation}
where $\psi(\vec{r}, p, t)$ represents the CR density per unit of total particle momentum $p$ at position $\vec{r}$, $Q(\vec{r}, p, t)$ describes the source term, $D_{x x}$ denotes the spatial diffusion coefficient, $\vec{V}_c$ is the convection velocity and $\tau_f$ and $\tau_r$ are the timescales for loss by fragmentation and radioactive decay, respectively. The CR diffusion depends on the distribution of CR sources $f(r,z)$, and the diffusion coefficient is described as\citep{2016ApJ...819...54G,2018PhRvD..97f3008G} 

\begin{equation}
D_{x x}(r, z, \mathcal{R})=D_0 F(r, z) \beta^\eta\left(\frac{\mathcal{R}}{\mathcal{R}_0}\right)^{\delta_0 F(r, z)},
\end{equation}
where $\delta_0 F(r, z)$ describes the turbulent characteristics of the local medium environment and $D_0 F(r, z)$ represents the normalization factor of the diffusion coefficient at the reference rigidity.
\begin{equation}
    F(r, z)= \begin{cases}g(r, z)+[1-g(r, z)]\left(\frac{z}{\xi z_0}\right)^n, & |z| \leq \xi z_0 \\ 1, & |z|>\xi z_0\end{cases},
\end{equation}
here, $\xi z_0$ denotes the half-thickness of the Galactic halo, and $g(r, z)=N_m /[1+f(r, z)]$, where $N_m$ is the normalization factor. The parameter $n$ is used to describe the smoothness between the inner and outer halos, and the source distribution $f(r,z)$ is a cylindrically symmetric continuous distribution\citep{1996A&AS..120C.437C,1998ApJ...509..212S,1998ApJ...504..761C}, 
\begin{equation}
f(r, z)=\left(\frac{r}{r_{\odot}}\right)^{1.25} \exp \left[-\frac{3.87\left(r-r_{\odot}\right)}{r_{\odot}}\right] \exp \left(-\frac{|z|}{z_s}\right),
\end{equation}
Where $r_{\odot}$ = 8.5 kpc and $z_s$ = 0.2 kpc. The Dpp describes the re-acceleration process of particles during propagation, and its coupling relationship with the spatial diffusion coefficient Dxx is given by:
\begin{equation}
D_{\mathrm{pp}} D_{\mathrm{xx}}=\frac{4 p^2 v_A^2}{3 \delta\left(4-\delta^2\right)(4-\delta)w}
\end{equation}
where $\mathrm{v_A}$ is the Alfvén speed, and w is the ratio of
magneto-hydrodynamic wave energy density to the magnetic field energy density, which can be fixed to 1.

\section{Results and Discussion} \label{sec:result}
\begin{figure}[t]%[htbp]
\centering
\includegraphics[width=0.95\linewidth]{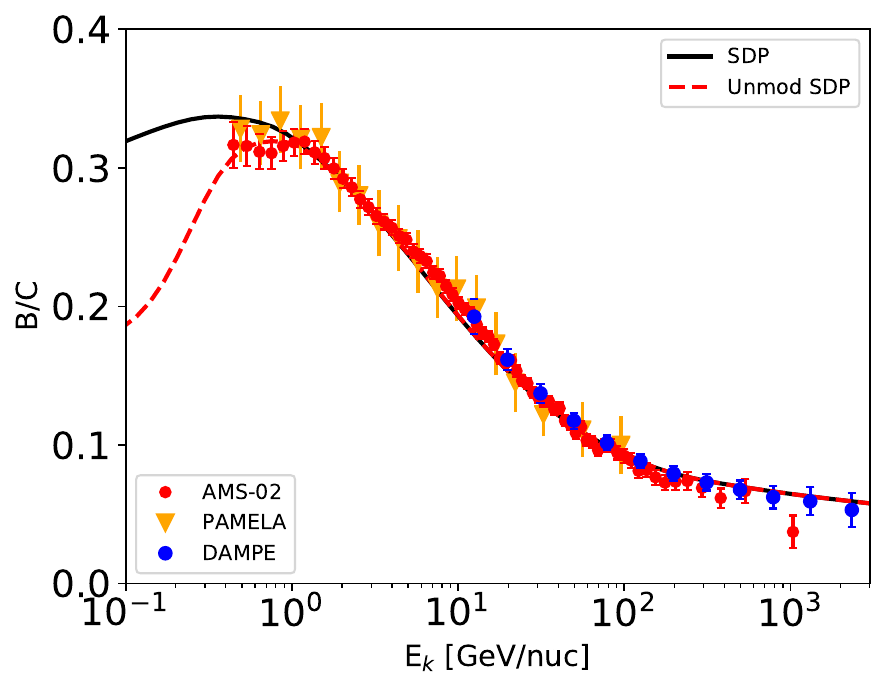}
\caption{A comparison of the B/C ratio calculated using the CR SDP model with observational data from AMS-02 \citep{2017PhRvL.119y1101A}, PAMELA \citep{2014ApJ...791...93A}, and DAMPE \citep{2022SciBu..67.2162D}. The red dashed line indicates the spectrum calculated without considering solar modulation. In this study, the solar modulation potential is consistently assumed to be 550 MeV.}
\label{BC}
\end{figure}

\begin{figure}[h]%[htbp]
    \centering
    \includegraphics[width=0.95\linewidth]{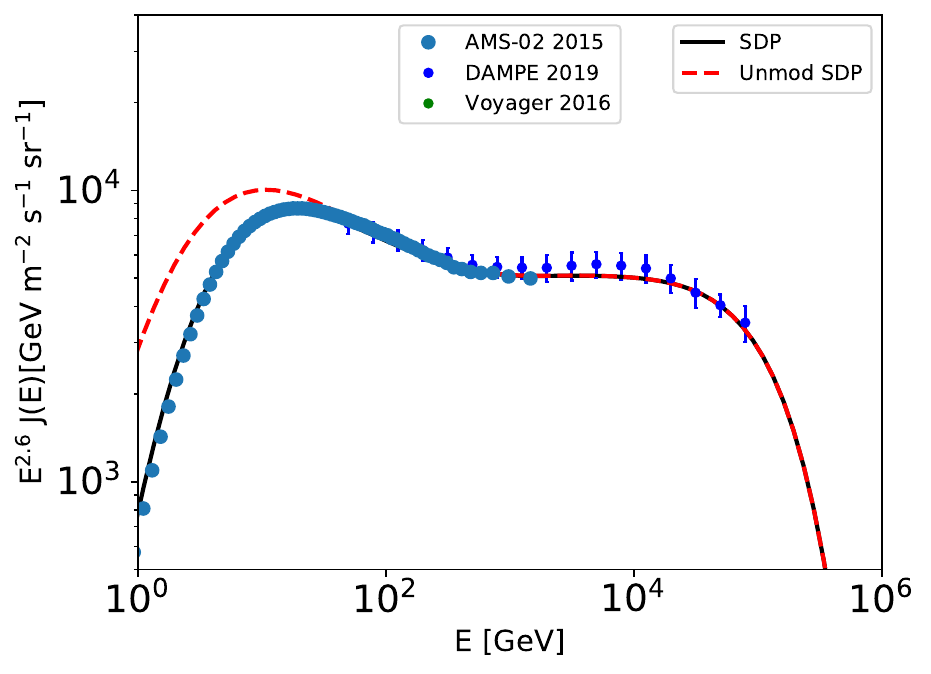}
    \caption{The CR proton spectrum is calculated using the CR SDP model and compared with observational data from AMS-02 \citep{2015PhRvL.114q1103A} and DAMPE \citep{2019SciA....5.3793A}. The red dashed line shows the spectrum without including solar modulation effects.}
    \label{proton}
    \end{figure}

To accurately estimate the gamma-ray emission generated by the background cosmic ray ``sea" near the Galactic center, we use the latest BC ratio and cosmic ray proton observation data to constrain the parameters of the propagation model. This allows us to calculate the spatial distribution of the background cosmic ray particles and, in turn, derive the diffuse gamma-ray spectrum for the region surrounding the Galactic center. As shown in Figure \ref{BC} and \ref{proton}, our calculations based on the spatially dependent cosmic ray propagation model successfully reproduce the observed cosmic ray data, with the corresponding model parameters listed in Table \ref{tab1}.

\begin{figure*}[t]%[htbp]
    \centering
    \includegraphics[width=0.95\linewidth]{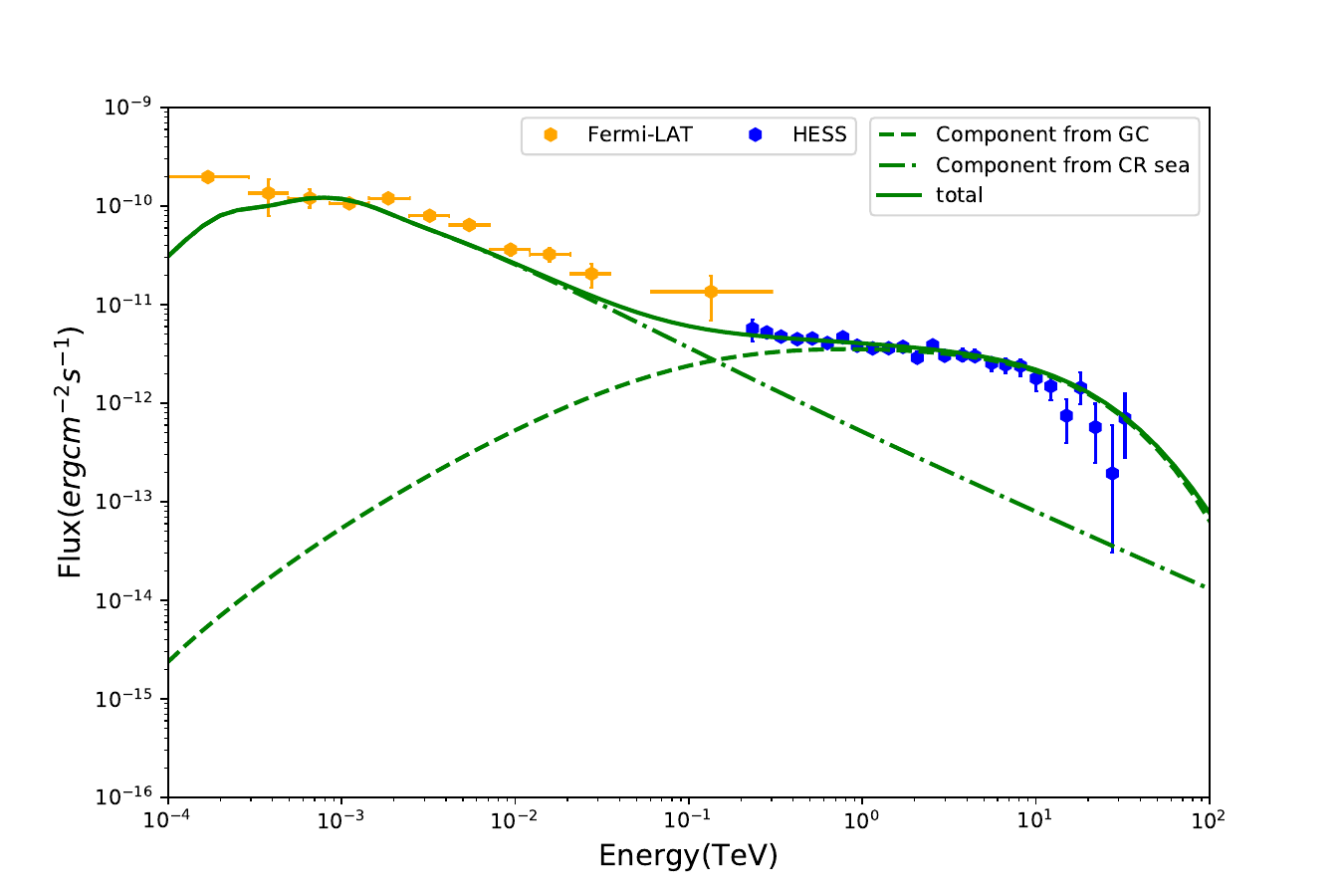}
    \caption{Combined Fermi (green points)\citep{2011ApJ...726...60C} and HESS (blue points)\citep{2009A&A...503..817A} explained by the superposition (green solid line) of a proton flare duration occurring 300 years ago (green dashed line) and a constant component that were generated by the background ``sea" cosmic rays trapped in the dense region near the Galactic center (green dashed-dotted line).}
    \label{GC_SED_figure}
    \end{figure*}

However, the gamma rays produced by cosmic ray protons through pp interactions not only depend on the injection and propagation of cosmic rays but also on the density of the surrounding gas. Observations suggest that a complex environment may exist around the Galactic center, with the presence of dense molecular clouds. Therefore, to better represent the true environment near the Galactic center, we assume that the black hole at the center is surrounded by dense gas, with a density of about $\rm 1000 cm^{-3}$ within a radius of approximately 3 pc. Using the GALPROP package\citep{1998ApJ...509..212S,2022ApJS..262...30P}, we then calculate the gamma-ray spectrum for a region centered on the Galactic center with a 3 pc radius, as shown in Figure \ref{GC_SED_figure}.

\begin{table}[t]%[htbp]
\footnotesize
\centering
\caption{Parameters of the SDP model.}
\label{tab1}
\tabcolsep 3.5pt
\begin{tabular*}{0.47\textwidth}{lcccccc}    
\hline \hline
$D_0{ }\left[\mathrm{cm}^{2} \mathrm{~s}^{-1}\right]$ & $\delta_0$ & $N_m$ & $\xi$ & $\mathrm{n}$ & $v_A\left[\mathrm{~km} \mathrm{~s}^{-1}\right]$ & $z_0[\mathrm{kpc}]$ \\
\hline $4.5 \times 10^{28}$ & 0.64 & 0.24 & 0.1 & 4.0 & 6 & 4.5 \\
\hline 
\end{tabular*}
\end{table}

The Galactic Center is a point-like astrophysical object, and there is substantial evidence indicating that it underwent a flare activity several hundred years ago. Consistent with previous research, we assume that cosmic ray protons were accelerated during the Galactic Center's flare activity several hundred years ago and were instantaneously injected into the dense environment approximately 3 parsecs around the Galactic Center, with an injection parameters of $\gamma=2.0$ and $\rm Q_0=8\times10^{39}ergs^{-1}$. To maintain consistency with the algorithm for the background ``sea" component mentioned above, we also utilize the galprop package to solve the time-dependent diffusion of cosmic ray protons injected from the Galactic Center and calculate the gamma-ray spectrum of the region of interest to us.
The resulting gamma-ray spectrum is shown in Figure \ref{GC_SED_figure}. As can be seen, the component from the Galactic center point source and the background ``sea" cosmic rays trapped in the dense region near the Galactic center combine well to reproduce the observational data from Fermi-LAT and HESS. On one hand, from a physical model perspective, the local source (Galactic center) dominates the high-energy (TeV) radiation spectrum, while the low-energy (GeV) radiation is dominated by interactions between the background ``sea" cosmic rays and the dense environmental gas near the Galactic center. Therefore, we believe that the low-energy component is a stable radiation component, while the Galactic center activity has a more significant impact on high-energy radiation. This is consistent with the spatially dependent propagation model that successfully explains the cosmic-ray observational spectrum. On the other hand, from an observational perspective, the combination of the limited resolution of Fermi-LAT and the great distance from the Galactic center to Earth may not be enough to completely subtract the diffuse gamma-ray background of the Milky Way.

In fact, our model uses two components to explain the data observed by Fermi-LAT and HESS, similar to previous work. For example, the light hadron mixing model proposed by the \cite{2013JPhG...40f5201G} suggests that the Fermi-LAT data is mainly produced by electrons accelerated by Galactic center activity interacting with the background radiation field via inverse Compton scattering, while the HESS data is formed by protons accelerated by the Galactic center interacting with the ambient gas. Some also argue that the different power-law spectra of the Galactic center in the GeV and TeV energy ranges are due to the acceleration of cosmic-ray protons at different times in the Galactic center, generating different components in the nearby region\citep{2011ApJ...726...60C}. However, compared to previous models, the most innovative aspect of our model is that the parent particles responsible for generating GeV gamma-ray radiation in our model originate from the background ``sea". While it is not yet possible to definitively identify the origin of the double power-law structure formed by the combined Fermi-LAT and HESS observational data, the model presented here, which fits these observational data well, may be generally applicable to most cosmic-ray halo sources in the Milky Way. It is expected that sources not observable in the GeV range may be detected in the TeV range. Future high-resolution observations of Galactic cosmic-ray sources will further validate our model.

\section{Conclusion and Summary} \label{sec:conclusion}
We propose a model based on the spatially dependent propagation of Galactic cosmic rays, which could be widely applicable to Galactic halo sources, and use this model to provide a possible explanation for the double power-law structure observed in the Galactic center’s GeV and TeV energy spectra. In this model, the high-energy radiation from the cosmic ray halo primarily originates from cosmic ray particles accelerated by point sources in its central region, while the low-energy radiation is due to the halo’s capture of background ``sea" cosmic ray particles. The radiation produced by these two components with different origins together forms the observed double power-law structure. The former cosmic rays are accelerated by point sources within the halo, depending on the properties of the local source, and exhibit free, spatial, and temporal dependence; while the latter cosmic ray particles are accelerated and propagated from distant cosmic ray sources, exhibiting stable and uniform characteristics. When the Galactic center is considered as a gamma-ray source surrounded by dense gas in the halo, the model calculations successfully reproduce the observational data from Fermi-LAT and HESS. Therefore, the observational data from the Galactic center not only support our model but also consistently support the spatially dependent cosmic ray propagation model. Based on our results, we believe that high-energy phenomena in the local region of the Milky Way (including the distribution of cosmic rays and diffuse gamma-rays) are dominated by local sources. We expect that some Galactic extended sources that are not detectable in the GeV range may be observable in the TeV range. We hope that future observations of more halo sources will further test our model.

\section*{Acknowledgements}
This work is supported in China by National Key R$\&$D program of China under the grant 2024YFA1611402, and supported by the National Natural Science Foundation of China (12275279,12375103).

\bibliographystyle{aasjournal}
\bibliography{ref_v2}{}

\begin{thebibliography}{}
\expandafter\ifx\csname natexlab\endcsname\relax\def\natexlab#1{#1}\fi
\providecommand{\url}[1]{\href{#1}{#1}}
\providecommand{\dodoi}[1]{doi:~\href{http://doi.org/#1}{\nolinkurl{#1}}}
\providecommand{\doeprint}[1]{\href{http://ascl.net/#1}{\nolinkurl{http://ascl.net/#1}}}
\providecommand{\doarXiv}[1]{\href{https://arxiv.org/abs/#1}{\nolinkurl{https://arxiv.org/abs/#1}}}

\bibitem[{{Abdollahi} {et~al.}(2017){Abdollahi}, {Ackermann}, {Ajello}, {Albert}, {Atwood}, {Baldini}, {Barbiellini}, {Bellazzini}, {Bissaldi}, {Bloom}, {Bonino}, {Bottacini}, {Brandt}, {Bruel}, {Buson}, {Caragiulo}, {Cavazzuti}, {Chekhtman}, {Ciprini}, {Costanza}, {Cuoco}, {Cutini}, {D'Ammando}, {de Palma}, {Desiante}, {Digel}, {Di Lalla}, {Di Mauro}, {Di Venere}, {Donaggio}, {Drell}, {Favuzzi}, {Focke}, {Fukazawa}, {Funk}, {Fusco}, {Gargano}, {Gasparrini}, {Giglietto}, {Giordano}, {Giroletti}, {Green}, {Guiriec}, {Harding}, {Jogler}, {J{\'o}hannesson}, {Kamae}, {Kuss}, {Larsson}, {Latronico}, {Li}, {Longo}, {Loparco}, {Lubrano}, {Magill}, {Malyshev}, {Manfreda}, {Mazziotta}, {Meehan}, {Michelson}, {Mitthumsiri}, {Mizuno}, {Moiseev}, {Monzani}, {Morselli}, {Negro}, {Nuss}, {Ohsugi}, {Omodei}, {Paneque}, {Perkins}, {Pesce-Rollins}, {Piron}, {Pivato}, {Principe}, {Rain{\`o}}, {Rando}, {Razzano}, {Reimer}, {Reimer}, {Sgr{\`o}}, {Simone}, {Siskind}, {Spada}, {Spandre}, {Spinelli}, {Strong}, {Tajima}, {Thayer},
  {Torres}, {Troja}, {Vandenbroucke}, {Zaharijas}, {Zimmer}, \& {Fermi-LAT Collaboration}}]{2017PhRvL.118i1103A}
{Abdollahi}, S., {Ackermann}, M., {Ajello}, M., {et~al.} 2017, \prl, 118, 091103, \dodoi{10.1103/PhysRevLett.118.091103}

\bibitem[{{Acero} {et~al.}(2016){Acero}, {Ackermann}, {Ajello}, {Albert}, {Baldini}, {Ballet}, {Barbiellini}, {Bastieri}, {Bellazzini}, {Bissaldi}, {Bloom}, {Bonino}, {Bottacini}, {Brandt}, {Bregeon}, {Bruel}, {Buehler}, {Buson}, {Caliandro}, {Cameron}, {Caragiulo}, {Caraveo}, {Casandjian}, {Cavazzuti}, {Cecchi}, {Charles}, {Chekhtman}, {Chiang}, {Chiaro}, {Ciprini}, {Claus}, {Cohen-Tanugi}, {Conrad}, {Cuoco}, {Cutini}, {D'Ammando}, {de Angelis}, {de Palma}, {Desiante}, {Digel}, {Di Venere}, {Drell}, {Favuzzi}, {Fegan}, {Ferrara}, {Focke}, {Franckowiak}, {Funk}, {Fusco}, {Gargano}, {Gasparrini}, {Giglietto}, {Giordano}, {Giroletti}, {Glanzman}, {Godfrey}, {Grenier}, {Guiriec}, {Hadasch}, {Harding}, {Hayashi}, {Hays}, {Hewitt}, {Hill}, {Horan}, {Hou}, {Jogler}, {J{\'o}hannesson}, {Kamae}, {Kuss}, {Landriu}, {Larsson}, {Latronico}, {Li}, {Li}, {Longo}, {Loparco}, {Lovellette}, {Lubrano}, {Maldera}, {Malyshev}, {Manfreda}, {Martin}, {Mayer}, {Mazziotta}, {McEnery}, {Michelson}, {Mirabal}, {Mizuno}, {Monzani},
  {Morselli}, {Nuss}, {Ohsugi}, {Omodei}, {Orienti}, {Orlando}, {Ormes}, {Paneque}, {Pesce-Rollins}, {Piron}, {Pivato}, {Rain{\`o}}, {Rando}, {Razzano}, {Razzaque}, {Reimer}, {Reimer}, {Remy}, {Renault}, {S{\'a}nchez-Conde}, {Schaal}, {Schulz}, {Sgr{\`o}}, {Siskind}, {Spada}, {Spandre}, {Spinelli}, {Strong}, {Suson}, {Tajima}, {Takahashi}, {Thayer}, {Thompson}, {Tibaldo}, {Tinivella}, {Torres}, {Tosti}, {Troja}, {Vianello}, {Werner}, {Wood}, {Wood}, {Zaharijas}, \& {Zimmer}}]{2016ApJS..223...26A}
{Acero}, F., {Ackermann}, M., {Ajello}, M., {et~al.} 2016, \apjs, 223, 26, \dodoi{10.3847/0067-0049/223/2/26}

\bibitem[{{Adriani} {et~al.}(2009){Adriani}, {Barbarino}, {Bazilevskaya}, {Bellotti}, {Boezio}, {Bogomolov}, {Bonechi}, {Bongi}, {Bonvicini}, {Bottai}, {Bruno}, {Cafagna}, {Campana}, {Carlson}, {Casolino}, {Castellini}, {de Pascale}, {de Rosa}, {de Simone}, {di Felice}, {Galper}, {Grishantseva}, {Hofverberg}, {Koldashov}, {Krutkov}, {Kvashnin}, {Leonov}, {Malvezzi}, {Marcelli}, {Menn}, {Mikhailov}, {Mocchiutti}, {Orsi}, {Osteria}, {Papini}, {Pearce}, {Picozza}, {Ricci}, {Ricciarini}, {Simon}, {Sparvoli}, {Spillantini}, {Stozhkov}, {Vacchi}, {Vannuccini}, {Vasilyev}, {Voronov}, {Yurkin}, {Zampa}, {Zampa}, \& {Zverev}}]{2009Natur.458..607A}
{Adriani}, O., {Barbarino}, G.~C., {Bazilevskaya}, G.~A., {et~al.} 2009, \nat, 458, 607, \dodoi{10.1038/nature07942}

\bibitem[{{Adriani} {et~al.}(2011){Adriani}, {Barbarino}, {Bazilevskaya}, {Bellotti}, {Boezio}, {Bogomolov}, {Bonechi}, {Bongi}, {Bonvicini}, {Borisov}, {Bottai}, {Bruno}, {Cafagna}, {Campana}, {Carbone}, {Carlson}, {Casolino}, {Castellini}, {Consiglio}, {De Pascale}, {De Santis}, {De Simone}, {Di Felice}, {Galper}, {Gillard}, {Grishantseva}, {Jerse}, {Karelin}, {Koldashov}, {Krutkov}, {Kvashnin}, {Leonov}, {Malakhov}, {Malvezzi}, {Marcelli}, {Mayorov}, {Menn}, {Mikhailov}, {Mocchiutti}, {Monaco}, {Mori}, {Nikonov}, {Osteria}, {Palma}, {Papini}, {Pearce}, {Picozza}, {Pizzolotto}, {Ricci}, {Ricciarini}, {Rossetto}, {Sarkar}, {Simon}, {Sparvoli}, {Spillantini}, {Stozhkov}, {Vacchi}, {Vannuccini}, {Vasilyev}, {Voronov}, {Yurkin}, {Wu}, {Zampa}, {Zampa}, \& {Zverev}}]{2011Sci...332...69A}
---. 2011, Science, 332, 69, \dodoi{10.1126/science.1199172}

\bibitem[{{Adriani} {et~al.}(2014){Adriani}, {Barbarino}, {Bazilevskaya}, {Bellotti}, {Boezio}, {Bogomolov}, {Bongi}, {Bonvicini}, {Bottai}, {Bruno}, {Cafagna}, {Campana}, {Carbone}, {Carlson}, {Casolino}, {Castellini}, {Danilchenko}, {De Donato}, {De Santis}, {De Simone}, {Di Felice}, {Formato}, {Galper}, {Karelin}, {Koldashov}, {Koldobskiy}, {Krutkov}, {Kvashnin}, {Leonov}, {Malakhov}, {Marcelli}, {Martucci}, {Mayorov}, {Menn}, {Merg{\'e}}, {Mikhailov}, {Mocchiutti}, {Monaco}, {Mori}, {Munini}, {Osteria}, {Palma}, {Panico}, {Papini}, {Pearce}, {Picozza}, {Pizzolotto}, {Ricci}, {Ricciarini}, {Rossetto}, {Sarkar}, {Scotti}, {Simon}, {Sparvoli}, {Spillantini}, {Stozhkov}, {Vacchi}, {Vannuccini}, {Vasilyev}, {Voronov}, {Yurkin}, {Zampa}, {Zampa}, \& {Zverev}}]{2014ApJ...791...93A}
---. 2014, \apj, 791, 93, \dodoi{10.1088/0004-637X/791/2/93}

\bibitem[{{Aguilar} {et~al.}(2015){Aguilar}, {Aisa}, {Alpat}, {Alvino}, {Ambrosi}, {Andeen}, {Arruda}, {Attig}, {Azzarello}, {Bachlechner}, {Barao}, {Barrau}, {Barrin}, {Bartoloni}, {Basara}, {Battarbee}, {Battiston}, {Bazo}, {Becker}, {Behlmann}, {Beischer}, {Berdugo}, {Bertucci}, {Bigongiari}, {Bindi}, {Bizzaglia}, {Bizzarri}, {Boella}, {de Boer}, {Bollweg}, {Bonnivard}, {Borgia}, {Borsini}, {Boschini}, {Bourquin}, {Burger}, {Cadoux}, {Cai}, {Capell}, {Caroff}, {Casaus}, {Cascioli}, {Castellini}, {Cernuda}, {Cerreta}, {Cervelli}, {Chae}, {Chang}, {Chen}, {Chen}, {Cheng}, {Chen}, {Cheng}, {Chou}, {Choumilov}, {Choutko}, {Chung}, {Clark}, {Clavero}, {Coignet}, {Consolandi}, {Contin}, {Corti}, {Gil}, {Coste}, {Creus}, {Crispoltoni}, {Cui}, {Dai}, {Delgado}, {Della Torre}, {Demirk{\"o}z}, {Derome}, {Di Falco}, {Di Masso}, {Dimiccoli}, {D{\'\i}az}, {von Doetinchem}, {Donnini}, {Du}, {Duranti}, {D'Urso}, {Eline}, {Eppling}, {Eronen}, {Fan}, {Farnesini}, {Feng}, {Fiandrini}, {Fiasson}, {Finch}, {Fisher},
  {Galaktionov}, {Gallucci}, {Garc{\'\i}a}, {Garc{\'\i}a-L{\'o}pez}, {Gargiulo}, {Gast}, {Gebauer}, {Gervasi}, {Ghelfi}, {Gillard}, {Giovacchini}, {Goglov}, {Gong}, {Goy}, {Grabski}, {Grandi}, {Graziani}, {Guandalini}, {Guerri}, {Guo}, {Haas}, {Habiby}, {Haino}, {Han}, {He}, {Heil}, {Hoffman}, {Hsieh}, {Huang}, {Huh}, {Incagli}, {Ionica}, {Jang}, {Jinchi}, {Kanishev}, {Kim}, {Kim}, {Kirn}, {Kossakowski}, {Kounina}, {Kounine}, {Koutsenko}, {Krafczyk}, {La Vacca}, {Laudi}, {Laurenti}, {Lazzizzera}, {Lebedev}, {Lee}, {Lee}, {Leluc}, {Levi}, {Li}, {Li}, {Li}, {Li}, {Li}, {Li}, {Li}, {Li}, {Li}, {Lim}, {Lin}, {Lipari}, {Lippert}, {Liu}, {Liu}, {Lolli}, {Lomtadze}, {Lu}, {Lu}, {Lu}, {Luebelsmeyer}, {Luo}, {Lv}, {Majka}, {Ma{\~n}{\'a}}, {Mar{\'\i}n}, {Martin}, {Mart{\'\i}nez}, {Masi}, {Maurin}, {Menchaca-Rocha}, {Meng}, {Mo}, {Morescalchi}, {Mott}, {M{\"u}ller}, {Ni}, {Nikonov}, {Nozzoli}, {Nunes}, {Obermeier}, {Oliva}, {Orcinha}, {Palmonari}, {Palomares}, {Paniccia}, {Papi}, {Pauluzzi}, {Pedreschi}, {Pensotti},
  {Pereira}, {Picot-Clemente}, {Pilo}, {Piluso}, {Pizzolotto}, {Plyaskin}, {Pohl}, {Poireau}, {Postaci}, {Putze}, {Quadrani}, {Qi}, {Qin}, {Qu}, {R{\"a}ih{\"a}}, {Rancoita}, {Rapin}, {Ricol}, {Rodr{\'\i}guez}, {Rosier-Lees}, {Rozhkov}, {Rozza}, {Sagdeev}, {Sandweiss}, {Saouter}, {Sbarra}, {Schael}, {Schmidt}, {von Dratzig}, {Schwering}, {Scolieri}, {Seo}, {Shan}, {Shan}, {Shi}, {Shi}, {Shi}, {Siedenburg}, {Son}, {Spada}, {Spinella}, {Sun}, {Sun}, {Tacconi}, {Tang}, {Tang}, {Tang}, {Tao}, {Tescaro}, {Ting}, {Ting}, {Tomassetti}, {Torsti}, {T{\"u}rko{\v{g}}lu}, {Urban}, {Vagelli}, {Valente}, {Vannini}, {Valtonen}, {Vaurynovich}, {Vecchi}, {Velasco}, {Vialle}, {Vitale}, {Vitillo}, {Wang}, {Wang}, {Wang}, {Wang}, {Wang}, {Wang}, {Weng}, {Whitman}, {Wienkenh{\"o}ver}, {Wu}, {Wu}, {Xia}, {Xie}, {Xie}, {Xiong}, {Xin}, {Xu}, {Xu}, {Yan}, {Yang}, {Yang}, {Ye}, {Yi}, {Yu}, {Yu}, {Zeissler}, {Zhang}, {Zhang}, {Zhang}, {Zhang}, {Zheng}, {Zhuang}, {Zhukov}, {Zichichi}, {Zimmermann}, {Zuccon}, {Zurbach}, \& {AMS
  Collaboration}}]{2015PhRvL.114q1103A}
{Aguilar}, M., {Aisa}, D., {Alpat}, B., {et~al.} 2015, \prl, 114, 171103, \dodoi{10.1103/PhysRevLett.114.171103}

\bibitem[{{Aguilar} {et~al.}(2016){Aguilar}, {Ali Cavasonza}, {Alpat}, {Ambrosi}, {Arruda}, {Attig}, {Aupetit}, {Azzarello}, {Bachlechner}, {Barao}, {Barrau}, {Barrin}, {Bartoloni}, {Basara}, {Ba{\c{s}}e{\c{C}}{\textsection}mez-du Pree}, {Battarbee}, {Battiston}, {Bazo}, {Becker}, {Behlmann}, {Beischer}, {Berdugo}, {Bertucci}, {Bindi}, {Boella}, {de Boer}, {Bollweg}, {Bonnivard}, {Borgia}, {Boschini}, {Bourquin}, {Bueno}, {Burger}, {Cadoux}, {Cai}, {Capell}, {Caroff}, {Casaus}, {Castellini}, {Cernuda}, {Cervelli}, {Chae}, {Chang}, {Chen}, {Chen}, {Chen}, {Cheng}, {Chou}, {Choumilov}, {Choutko}, {Chung}, {Clark}, {Clavero}, {Coignet}, {Consolandi}, {Contin}, {Corti}, {Coste}, {Creus}, {Crispoltoni}, {Cui}, {Dai}, {Delgado}, {Della Torre}, {Demirk{\"o}z}, {Derome}, {Di Falco}, {Dimiccoli}, {D{\'\i}az}, {von Doetinchem}, {Dong}, {Donnini}, {Duranti}, {D'Urso}, {Egorov}, {Eline}, {Eronen}, {Feng}, {Fiandrini}, {Finch}, {Fisher}, {Formato}, {Galaktionov}, {Gallucci}, {Garc{\'\i}a}, {Garc{\'\i}a-L{\'o}pez},
  {Gargiulo}, {Gast}, {Gebauer}, {Gervasi}, {Ghelfi}, {Giovacchini}, {Goglov}, {G{\'o}mez-Coral}, {Gong}, {Goy}, {Grabski}, {Grandi}, {Graziani}, {Guerri}, {Guo}, {Habiby}, {Haino}, {Han}, {He}, {Heil}, {Hoffman}, {Hsieh}, {Huang}, {Huang}, {Huh}, {Incagli}, {Ionica}, {Jang}, {Jinchi}, {Kang}, {Kanishev}, {Kim}, {Kim}, {Kirn}, {Konak}, {Kounina}, {Kounine}, {Koutsenko}, {Krafczyk}, {La Vacca}, {Laudi}, {Laurenti}, {Lazzizzera}, {Lebedev}, {Lee}, {Lee}, {Leluc}, {Li}, {Li}, {Li}, {Li}, {Li}, {Li}, {Li}, {Li}, {Lim}, {Lin}, {Lipari}, {Lippert}, {Liu}, {Liu}, {Lu}, {Lu}, {Luebelsmeyer}, {Luo}, {Luo}, {Lv}, {Majka}, {Ma{\~n}{\'a}}, {Mar{\'\i}n}, {Martin}, {Mart{\'\i}nez}, {Masi}, {Maurin}, {Menchaca-Rocha}, {Meng}, {Mo}, {Morescalchi}, {Mott}, {Nelson}, {Ni}, {Nikonov}, {Nozzoli}, {Nunes}, {Oliva}, {Orcinha}, {Palmonari}, {Palomares}, {Paniccia}, {Pauluzzi}, {Pensotti}, {Pereira}, {Picot-Clemente}, {Pilo}, {Pizzolotto}, {Plyaskin}, {Pohl}, {Poireau}, {Putze}, {Quadrani}, {Qi}, {Qin}, {Qu}, {R{\"a}ih{\"a}},
  {Rancoita}, {Rapin}, {Ricol}, {Rodr{\'\i}guez}, {Rosier-Lees}, {Rozhkov}, {Rozza}, {Sagdeev}, {Sandweiss}, \& {Saouter}}]{2016PhRvL.117i1103A}
{Aguilar}, M., {Ali Cavasonza}, L., {Alpat}, B., {et~al.} 2016, \prl, 117, 091103, \dodoi{10.1103/PhysRevLett.117.091103}

\bibitem[{{Aguilar} {et~al.}(2017){Aguilar}, {Ali Cavasonza}, {Alpat}, {Ambrosi}, {Arruda}, {Attig}, {Aupetit}, {Azzarello}, {Bachlechner}, {Barao}, {Barrau}, {Barrin}, {Bartoloni}, {Basara}, {Ba{\c{s}}e{\v{g}}mez-du Pree}, {Battarbee}, {Battiston}, {Becker}, {Behlmann}, {Beischer}, {Berdugo}, {Bertucci}, {Bindel}, {Bindi}, {de Boer}, {Bollweg}, {Bonnivard}, {Borgia}, {Boschini}, {Bourquin}, {Bueno}, {Burger}, {Burger}, {Cadoux}, {Cai}, {Capell}, {Caroff}, {Casaus}, {Castellini}, {Cervelli}, {Chae}, {Chang}, {Chen}, {Chen}, {Chen}, {Cheng}, {Chou}, {Choumilov}, {Choutko}, {Chung}, {Clark}, {Clavero}, {Coignet}, {Consolandi}, {Contin}, {Corti}, {Creus}, {Crispoltoni}, {Cui}, {Dadzie}, {Dai}, {Datta}, {Delgado}, {Della Torre}, {Demakov}, {Demirk{\"o}z}, {Derome}, {Di Falco}, {Dimiccoli}, {D{\'\i}az}, {von Doetinchem}, {Dong}, {Donnini}, {Duranti}, {D'Urso}, {Egorov}, {Eline}, {Eronen}, {Feng}, {Fiandrini}, {Fisher}, {Formato}, {Galaktionov}, {Gallucci}, {Garc{\'\i}a-L{\'o}pez}, {Gargiulo}, {Gast}, {Gebauer},
  {Gervasi}, {Ghelfi}, {Giovacchini}, {G{\'o}mez-Coral}, {Gong}, {Goy}, {Grabski}, {Grandi}, {Graziani}, {Guo}, {Haino}, {Han}, {He}, {Heil}, {Hoffman}, {Hsieh}, {Huang}, {Huang}, {Huh}, {Incagli}, {Ionica}, {Jang}, {Jia}, {Jinchi}, {Kang}, {Kanishev}, {Khiali}, {Kim}, {Kim}, {Kirn}, {Konak}, {Kounina}, {Kounine}, {Koutsenko}, {Kulemzin}, {La Vacca}, {Laudi}, {Laurenti}, {Lazzizzera}, {Lebedev}, {Lee}, {Lee}, {Leluc}, {Li}, {Li}, {Li}, {Li}, {Li}, {Li}, {Li}, {Lim}, {Lin}, {Lipari}, {Lippert}, {Liu}, {Liu}, {Lordello}, {Lu}, {Lu}, {Luebelsmeyer}, {Luo}, {Luo}, {Lyu}, {Machate}, {Ma{\~n}{\'a}}, {Mar{\'\i}n}, {Martin}, {Mart{\'\i}nez}, {Masi}, {Maurin}, {Menchaca-Rocha}, {Meng}, {Mikuni}, {Mo}, {Mott}, {Nelson}, {Ni}, {Nikonov}, {Nozzoli}, {Oliva}, {Orcinha}, {Palmonari}, {Palomares}, {Paniccia}, {Pauluzzi}, {Pensotti}, {Perrina}, {Phan}, {Picot-Clemente}, {Pilo}, {Pizzolotto}, {Plyaskin}, {Pohl}, {Poireau}, {Quadrani}, {Qi}, {Qin}, {Qu}, {R{\"a}ih{\"a}}, {Rancoita}, {Rapin}, {Ricol}, {Rosier-Lees}, {Rozhkov},
  {Rozza}, {Sagdeev}, {Schael}, {Schmidt}, {Schulz von Dratzig}, {Schwering}, {Seo}, {Shan}, {Shi}, {Siedenburg}, {Son}, {Song}, {Tacconi}, {Tang}, {Tang}, {Tescaro}, {Ting}, {Ting}, {Tomassetti}, {Torsti}, {T{\"u}rko{\v{g}}lu}, {Urban}, {Vagelli}, {Valente}, {Valtonen}, {V{\'a}zquez Acosta}, {Vecchi}, {Velasco}, {Vialle}, {Vitale}, {Vitillo}, {Wang}, {Wang}, {Wang}, {Wang}, {Wang}, {Wang}, {Wei}, {Weng}, {Whitman}, {Wu}, {Wu}, {Xiong}, {Xu}, {Yan}, {Yang}, {Yang}, {Yang}, {Yi}, {Yu}, {Yu}, {Zannoni}, {Zeissler}, {Zhang}, {Zhang}, {Zhang}, {Zhang}, {Zhang}, {Zhang}, {Zheng}, {Zhuang}, {Zhukov}, {Zichichi}, {Zimmermann}, {Zuccon}, \& {AMS Collaboration}}]{2017PhRvL.119y1101A}
---. 2017, \prl, 119, 251101, \dodoi{10.1103/PhysRevLett.119.251101}

\bibitem[{{Aguilar} {et~al.}(2018){Aguilar}, {Ali Cavasonza}, {Ambrosi}, {Arruda}, {Attig}, {Aupetit}, {Azzarello}, {Bachlechner}, {Barao}, {Barrau}, {Barrin}, {Bartoloni}, {Basara}, {Ba{\c{s}}e{\v{g}}mez-du Pree}, {Battarbee}, {Battiston}, {Becker}, {Behlmann}, {Beischer}, {Berdugo}, {Bertucci}, {Bindel}, {Bindi}, {de Boer}, {Bollweg}, {Bonnivard}, {Borgia}, {Boschini}, {Bourquin}, {Bueno}, {Burger}, {Burger}, {Cadoux}, {Cai}, {Capell}, {Caroff}, {Casaus}, {Castellini}, {Cervelli}, {Chae}, {Chang}, {Chen}, {Chen}, {Chen}, {Cheng}, {Chou}, {Choumilov}, {Choutko}, {Chung}, {Clark}, {Clavero}, {Coignet}, {Consolandi}, {Contin}, {Corti}, {Creus}, {Crispoltoni}, {Cui}, {Dadzie}, {Dai}, {Datta}, {Delgado}, {Della Torre}, {Demirk{\"o}z}, {Derome}, {Di Falco}, {Dimiccoli}, {D{\'\i}az}, {von Doetinchem}, {Dong}, {Donnini}, {Duranti}, {D'Urso}, {Egorov}, {Eline}, {Eronen}, {Feng}, {Fiandrini}, {Fisher}, {Formato}, {Galaktionov}, {Gallucci}, {Garc{\'\i}a-L{\'o}pez}, {Gargiulo}, {Gast}, {Gebauer}, {Gervasi}, {Ghelfi},
  {Giovacchini}, {G{\'o}mez-Coral}, {Gong}, {Goy}, {Grabski}, {Grandi}, {Graziani}, {Guo}, {Haino}, {Han}, {He}, {Heil}, {Hsieh}, {Huang}, {Huang}, {Huh}, {Incagli}, {Ionica}, {Jang}, {Jia}, {Jinchi}, {Kang}, {Kanishev}, {Khiali}, {Kim}, {Kim}, {Kirn}, {Konak}, {Kounina}, {Kounine}, {Koutsenko}, {Kulemzin}, {La Vacca}, {Laudi}, {Laurenti}, {Lazzizzera}, {Lebedev}, {Lee}, {Lee}, {Leluc}, {Li}, {Li}, {Li}, {Li}, {Li}, {Li}, {Li}, {Lim}, {Lin}, {Lipari}, {Lippert}, {Liu}, {Liu}, {Lordello}, {Lu}, {Lu}, {Luebelsmeyer}, {Luo}, {Luo}, {Lyu}, {Machate}, {Ma{\~n}{\'a}}, {Mar{\'\i}n}, {Martin}, {Mart{\'\i}nez}, {Masi}, {Maurin}, {Menchaca-Rocha}, {Meng}, {Mikuni}, {Mo}, {Mott}, {Nelson}, {Ni}, {Nikonov}, {Nozzoli}, {Oliva}, {Orcinha}, {Palermo}, {Palmonari}, {Palomares}, {Paniccia}, {Pauluzzi}, {Pensotti}, {Perrina}, {Phan}, {Picot-Clemente}, {Pilo}, {Pizzolotto}, {Plyaskin}, {Pohl}, {Poireau}, {Quadrani}, {Qi}, {Qin}, {Qu}, {R{\"a}ih{\"a}}, {Rancoita}, {Rapin}, {Ricol}, {Rosier-Lees}, {Rozhkov}, {Rozza}, {Sagdeev},
  {Schael}, {Schmidt}, {Schulz von Dratzig}, {Schwering}, {Seo}, {Shan}, {Shi}, \& {Siedenburg}}]{2018PhRvL.120b1101A}
{Aguilar}, M., {Ali Cavasonza}, L., {Ambrosi}, G., {et~al.} 2018, \prl, 120, 021101, \dodoi{10.1103/PhysRevLett.120.021101}

\bibitem[{{Aguilar} {et~al.}(2021){Aguilar}, {Ali Cavasonza}, {Ambrosi}, {Arruda}, {Attig}, {Barao}, {Barrin}, {Bartoloni}, {Ba{\c{s}}e{\u{g}}mez-du Pree}, {Bates}, {Battiston}, {Behlmann}, {Beischer}, {Berdugo}, {Bertucci}, {Bindi}, {de Boer}, {Bollweg}, {Borgia}, {Boschini}, {Bourquin}, {Bueno}, {Burger}, {Burger}, {Burmeister}, {Cai}, {Capell}, {Casaus}, {Castellini}, {Cervelli}, {Chang}, {Chen}, {Chen}, {Chen}, {Cheng}, {Chou}, {Chouridou}, {Choutko}, {Chung}, {Clark}, {Coignet}, {Consolandi}, {Contin}, {Corti}, {Cui}, {Dadzie}, {Dai}, {Delgado}, {Della Torre}, {Demirk{\"o}z}, {Derome}, {Di Falco}, {Di Felice}, {D{\'\i}az}, {Dimiccoli}, {von Doetinchem}, {Dong}, {Donnini}, {Duranti}, {Egorov}, {Eline}, {Feng}, {Fiandrini}, {Fisher}, {Formato}, {Freeman}, {Galaktionov}, {G{\'a}mez}, {Garc{\'\i}a-L{\'o}pez}, {Gargiulo}, {Gast}, {Gebauer}, {Gervasi}, {Giovacchini}, {G{\'o}mez-Coral}, {Gong}, {Goy}, {Grabski}, {Grandi}, {Graziani}, {Guo}, {Haino}, {Han}, {Hashmani}, {He}, {Heber}, {Hsieh}, {Hu}, {Huang},
  {Hungerford}, {Incagli}, {Jang}, {Jia}, {Jinchi}, {Kanishev}, {Khiali}, {Kim}, {Kirn}, {Konyushikhin}, {Kounina}, {Kounine}, {Koutsenko}, {Kuhlman}, {Kulemzin}, {La Vacca}, {Laudi}, {Laurenti}, {Lazzizzera}, {Lebedev}, {Lee}, {Lee}, {Leluc}, {Li}, {Li}, {Li}, {Li}, {Li}, {Li}, {Light}, {Lin}, {Lippert}, {Liu}, {Lu}, {Lu}, {Luebelsmeyer}, {Luo}, {Lyu}, {Machate}, {Ma{\~n}{\'a}}, {Mar{\'\i}n}, {Marquardt}, {Martin}, {Mart{\'\i}nez}, {Masi}, {Maurin}, {Menchaca-Rocha}, {Meng}, {Mo}, {Molero}, {Mott}, {Mussolin}, {Ni}, {Nikonov}, {Nozzoli}, {Oliva}, {Orcinha}, {Palermo}, {Palmonari}, {Paniccia}, {Pashnin}, {Pauluzzi}, {Pensotti}, {Phan}, {Plyaskin}, {Pohl}, {Porter}, {Qi}, {Qin}, {Qu}, {Quadrani}, {Rancoita}, {Rapin}, {Reina Conde}, {Rosier-Lees}, {Rozhkov}, {Rozza}, {Sagdeev}, {Schael}, {Schmidt}, {Schulz von Dratzig}, {Schwering}, {Seo}, {Shan}, {Shi}, {Siedenburg}, {Solano}, {Song}, {Sonnabend}, {Sun}, {Sun}, {Tacconi}, {Tang}, {Tang}, {Tian}, {Ting}, {Ting}, {Tomassetti}, {Torsti}, {T{\"u}ys{\"u}z},
  {Urban}, {Usoskin}, {Vagelli}, {Vainio}, {Valente}, {Valtonen}, {V{\'a}zquez Acosta}, {Vecchi}, {Velasco}, {Vialle}, {Wang}, {Wang}, {Wang}, {Wang}, {Wang}, {Wang}, {Wei}, {Weng}, {Wu}, {Xiong}, {Xu}, {Yan}, {Yang}, {Yi}, {Yu}, {Yu}, {Zannoni}, {Zhang}, {Zhang}, {Zhang}, {Zhang}, {Zhang}, {Zhao}, {Zheng}, {Zhuang}, {Zhukov}, {Zichichi}, {Zimmermann}, {Zuccon}, \& {AMS Collaboration}}]{2021PhR...894....1A}
---. 2021, \physrep, 894, 1, \dodoi{10.1016/j.physrep.2020.09.003}

\bibitem[{{Aharonian} {et~al.}(2009){Aharonian}, {Akhperjanian}, {Anton}, {Barres de Almeida}, {Bazer-Bachi}, {Becherini}, {Behera}, {Bernl{\"o}hr}, {Boisson}, {Bochow}, {Borrel}, {Braun}, {Brion}, {Brucker}, {Brun}, {B{\"u}hler}, {Bulik}, {B{\"u}sching}, {Boutelier}, {Chadwick}, {Charbonnier}, {Chaves}, {Cheesebrough}, {Chounet}, {Clapson}, {Coignet}, {Dalton}, {Daniel}, {Davids}, {Degrange}, {Deil}, {Dickinson}, {Djannati-Ata{\"\i}}, {Domainko}, {O'C. Drury}, {Dubois}, {Dubus}, {Dyks}, {Dyrda}, {Egberts}, {Emmanoulopoulos}, {Espigat}, {Farnier}, {Feinstein}, {Fiasson}, {F{\"o}rster}, {Fontaine}, {F{\"u}{\ss}ling}, {Gabici}, {Gallant}, {G{\'e}rard}, {Giebels}, {Glicenstein}, {Gl{\"u}ck}, {Goret}, {Hauser}, {Hauser}, {Heinz}, {Heinzelmann}, {Henri}, {Hermann}, {Hinton}, {Hoffmann}, {Hofmann}, {Holleran}, {Hoppe}, {Horns}, {Jacholkowska}, {de Jager}, {Jung}, {Katarzy{\'n}ski}, {Katz}, {Kaufmann}, {Kendziorra}, {Kerschhaggl}, {Khangulyan}, {Kh{\'e}lifi}, {Keogh}, {Komin}, {Kosack}, {Lamanna}, {Lenain}, {Lohse},
  {Marandon}, {Martin}, {Martineau-Huynh}, {Marcowith}, {Maurin}, {McComb}, {Medina}, {Moderski}, {Moulin}, {Naumann-Godo}, {de Naurois}, {Nedbal}, {Nekrassov}, {Niemiec}, {Nolan}, {Ohm}, {Olive}, {de O{\~n}a Wilhelmi}, {Orford}, {Ostrowski}, {Panter}, {Paz Arribas}, {Pedaletti}, {Pelletier}, {Petrucci}, {Pita}, {P{\"u}hlhofer}, {Punch}, {Quirrenbach}, {Raubenheimer}, {Raue}, {Rayner}, {Renaud}, {Rieger}, {Ripken}, {Rob}, {Rolland}, {Rosier-Lees}, {Rowell}, {Rudak}, {Rulten}, {Ruppel}, {Sahakian}, {Santangelo}, {Schlickeiser}, {Sch{\"o}ck}, {Schr{\"o}der}, {Schwanke}, {Schwarzburg}, {Schwemmer}, {Shalchi}, {Skilton}, {Sol}, {Spangler}, {Stawarz}, {Steenkamp}, {Stegmann}, {Superina}, {Szostek}, {Tam}, {Tavernet}, {Terrier}, {Tibolla}, {van Eldik}, {Vasileiadis}, {Venter}, {Venter}, {Vialle}, {Vincent}, {Vivier}, {V{\"o}lk}, {Volpe}, {Wagner}, {Ward}, {Zdziarski}, \& {Zech}}]{2009A&A...503..817A}
{Aharonian}, F., {Akhperjanian}, A.~G., {Anton}, G., {et~al.} 2009, \aap, 503, 817, \dodoi{10.1051/0004-6361/200811569}

\bibitem[{{Ahn} {et~al.}(2010){Ahn}, {Allison}, {Bagliesi}, {Beatty}, {Bigongiari}, {Childers}, {Conklin}, {Coutu}, {DuVernois}, {Ganel}, {Han}, {Jeon}, {Kim}, {Lee}, {Lutz}, {Maestro}, {Malinin}, {Marrocchesi}, {Minnick}, {Mognet}, {Nam}, {Nam}, {Nutter}, {Park}, {Park}, {Seo}, {Sina}, {Wu}, {Yang}, {Yoon}, {Zei}, \& {Zinn}}]{2010ApJ...714L..89A}
{Ahn}, H.~S., {Allison}, P., {Bagliesi}, M.~G., {et~al.} 2010, \apjl, 714, L89, \dodoi{10.1088/2041-8205/714/1/L89}

\bibitem[{{An} {et~al.}(2019){An}, {Asfandiyarov}, {Azzarello}, {Bernardini}, {Bi}, {Cai}, {Chang}, {Chen}, {Chen}, {Chen}, {Chen}, {Cui}, {Cui}, {Dai}, {D'Amone}, {De Benedittis}, {De Mitri}, {Di Santo}, {Ding}, {Dong}, {Dong}, {Dong}, {Donvito}, {Droz}, {Duan}, {Duan}, {D'Urso}, {Fan}, {Fan}, {Fang}, {Feng}, {Feng}, {Fusco}, {Gallo}, {Gan}, {Gao}, {Gargano}, {Gong}, {Gong}, {Guo}, {Guo}, {Guo}, {Han}, {Hu}, {Huang}, {Huang}, {Huang}, {Ionica}, {Jiang}, {Jin}, {Kong}, {Lei}, {Li}, {Li}, {Li}, {Li}, {Li}, {Liang}, {Liang}, {Liao}, {Liu}, {Liu}, {Liu}, {Liu}, {Liu}, {Liu}, {Loparco}, {Luo}, {Ma}, {Ma}, {Ma}, {Ma}, {Ma}, {Marsella}, {Mazziotta}, {Mo}, {Niu}, {Pan}, {Peng}, {Peng}, {Qiao}, {Rao}, {Salinas}, {Shang}, {Shen}, {Shen}, {Shen}, {Song}, {Su}, {Su}, {Sun}, {Surdo}, {Teng}, {Tykhonov}, {Vitillo}, {Wang}, {Wang}, {Wang}, {Wang}, {Wang}, {Wang}, {Wang}, {Wang}, {Wang}, {Wang}, {Wang}, {Wang}, {Wang}, {Wei}, {Wei}, {Wei}, {Wen}, {Wu}, {Wu}, {Wu}, {Wu}, {Wu}, {Xi}, {Xia}, {Xu}, {Xu}, {Xu}, {Xu}, {Xue},
  {Yang}, {Yang}, {Yang}, {Yang}, {Yao}, {Yu}, {Yuan}, {Yue}, {Zang}, {Zhang}, {Zhang}, {Zhang}, {Zhang}, {Zhang}, {Zhang}, {Zhang}, {Zhang}, {Zhang}, {Zhang}, {Zhang}, {Zhang}, {Zhang}, {Zhao}, {Zhao}, {Zhao}, {Zhou}, {Zhou}, {Zhu}, {Zhu}, \& {Zimmer}}]{2019SciA....5.3793A}
{An}, Q., {Asfandiyarov}, R., {Azzarello}, P., {et~al.} 2019, Science Advances, 5, eaax3793, \dodoi{10.1126/sciadv.aax3793}

\bibitem[{{Case} \& {Bhattacharya}(1996)}]{1996A&AS..120C.437C}
{Case}, G., \& {Bhattacharya}, D. 1996, \aaps, 120, 437

\bibitem[{{Case} \& {Bhattacharya}(1998)}]{1998ApJ...504..761C}
{Case}, G.~L., \& {Bhattacharya}, D. 1998, \apj, 504, 761, \dodoi{10.1086/306089}

\bibitem[{{Chernyakova} {et~al.}(2011){Chernyakova}, {Malyshev}, {Aharonian}, {Crocker}, \& {Jones}}]{2011ApJ...726...60C}
{Chernyakova}, M., {Malyshev}, D., {Aharonian}, F.~A., {Crocker}, R.~M., \& {Jones}, D.~I. 2011, \apj, 726, 60, \dodoi{10.1088/0004-637X/726/2/60}

\bibitem[{{Dampe Collaboration}(2022)}]{2022SciBu..67.2162D}
{Dampe Collaboration}. 2022, Science Bulletin, 67, 2162, \dodoi{10.1016/j.scib.2022.10.002}

\bibitem[{{Fatuzzo} \& {Melia}(2012)}]{2012ApJ...757L..16F}
{Fatuzzo}, M., \& {Melia}, F. 2012, \apjl, 757, L16, \dodoi{10.1088/2041-8205/757/1/L16}

\bibitem[{{Guo} {et~al.}(2016){Guo}, {Tian}, \& {Jin}}]{2016ApJ...819...54G}
{Guo}, Y.-Q., {Tian}, Z., \& {Jin}, C. 2016, \apj, 819, 54, \dodoi{10.3847/0004-637X/819/1/54}

\bibitem[{{Guo} \& {Yuan}(2018)}]{2018PhRvD..97f3008G}
{Guo}, Y.-Q., \& {Yuan}, Q. 2018, \prd, 97, 063008, \dodoi{10.1103/PhysRevD.97.063008}

\bibitem[{{Guo} {et~al.}(2013){Guo}, {Yuan}, {Liu}, \& {Li}}]{2013JPhG...40f5201G}
{Guo}, Y.-Q., {Yuan}, Q., {Liu}, C., \& {Li}, A.-F. 2013, Journal of Physics G Nuclear Physics, 40, 065201, \dodoi{10.1088/0954-3899/40/6/065201}

\bibitem[{{J{\'o}hannesson} {et~al.}(2019){J{\'o}hannesson}, {Porter}, \& {Moskalenko}}]{2019ApJ...879...91J}
{J{\'o}hannesson}, G., {Porter}, T.~A., \& {Moskalenko}, I.~V. 2019, \apj, 879, 91, \dodoi{10.3847/1538-4357/ab258e}

\bibitem[{{Kusunose} \& {Takahara}(2012)}]{2012ApJ...748...34K}
{Kusunose}, M., \& {Takahara}, F. 2012, \apj, 748, 34, \dodoi{10.1088/0004-637X/748/1/34}

\bibitem[{{Linden} {et~al.}(2012){Linden}, {Lovegrove}, \& {Profumo}}]{2012ApJ...753...41L}
{Linden}, T., {Lovegrove}, E., \& {Profumo}, S. 2012, \apj, 753, 41, \dodoi{10.1088/0004-637X/753/1/41}

\bibitem[{{Nie} {et~al.}(2024){Nie}, {Qian}, {Guo}, \& {Liu}}]{2024ApJ...974..276N}
{Nie}, L., {Qian}, X.-L., {Guo}, Y.-Q., \& {Liu}, S.-M. 2024, \apj, 974, 276, \dodoi{10.3847/1538-4357/ad7eab}

\bibitem[{{Panov} {et~al.}(2007){Panov}, {Adams}, {Ahn}, {Batkov}, {Bashindzhagyan}, {Watts}, {Wefel}, {Wu}, {Ganel}, {Guzik}, {Gunashingha}, {Zatsepin}, {Isbert}, {Kim}, {Christl}, {Kouznetsov}, {Panasyuk}, {Seo}, {Sokolskaya}, {Chang}, {Schmidt}, \& {Fazely}}]{2007BRASP..71..494P}
{Panov}, A.~D., {Adams}, Jr., J.~H., {Ahn}, H.~S., {et~al.} 2007, Bulletin of the Russian Academy of Sciences, Physics, 71, 494, \dodoi{10.3103/S1062873807040168}

\bibitem[{{Panov} {et~al.}(2009){Panov}, {Adams}, {Ahn}, {Bashinzhagyan}, {Watts}, {Wefel}, {Wu}, {Ganel}, {Guzik}, {Zatsepin}, {Isbert}, {Kim}, {Christl}, {Kouznetsov}, {Panasyuk}, {Seo}, {Sokolskaya}, {Chang}, {Schmidt}, \& {Fazely}}]{2009BRASP..73..564P}
{Panov}, A.~D., {Adams}, J.~H., {Ahn}, H.~S., {et~al.} 2009, Bulletin of the Russian Academy of Sciences, Physics, 73, 564, \dodoi{10.3103/S1062873809050098}

\bibitem[{{Porter} {et~al.}(2022){Porter}, {J{\'o}hannesson}, \& {Moskalenko}}]{2022ApJS..262...30P}
{Porter}, T.~A., {J{\'o}hannesson}, G., \& {Moskalenko}, I.~V. 2022, \apjs, 262, 30, \dodoi{10.3847/1538-4365/ac80f6}

\bibitem[{{Strong} \& {Moskalenko}(1998)}]{1998ApJ...509..212S}
{Strong}, A.~W., \& {Moskalenko}, I.~V. 1998, \apj, 509, 212, \dodoi{10.1086/306470}

\bibitem[{{Yao} {et~al.}(2024){Yao}, {Dong}, {Guo}, \& {Yuan}}]{2024PhRvD.109f3001Y}
{Yao}, Y.-H., {Dong}, X.-L., {Guo}, Y.-Q., \& {Yuan}, Q. 2024, \prd, 109, 063001, \dodoi{10.1103/PhysRevD.109.063001}

\end{thebibliography}
\end{document}